\documentclass[preprint2]{aastex}

\begin{document}


\title{Spatial motion of the Magellanic Clouds.\\
Tidal models ruled out?}

\author{Adam R\accent23u\v{z}i\v{c}ka\altaffilmark{1,2}, Christian Theis\altaffilmark{2}, \& Jan Palou\v{s}\altaffilmark{1}}

\altaffiltext{1}{Astronomical Institute, Academy of Sciences of the Czech Republic, v.v.i.,
Bo\v{c}n\'{i} II 1401, 141\,31, Prague, adam.ruzicka@gmail.com}
\altaffiltext{2}{Institut f\"{u}r Astronomie der Universit\"{a}t Wien,
T\"urkenschanzstrasse 17, A-1180 Wien, Austria}

\begin{abstract}
Recently, Kallivayalil et al. derived new values of the proper motion for
the Large and Small Magellanic Clouds (LMC and SMC, respectively).
The spatial velocities of both Clouds are
unexpectedly higher than their previous values resulting from
agreement between the available
theoretical models of the Magellanic System and the observations of neutral hydrogen (H\,I)
associated with the LMC and the SMC.
Such proper motion estimates
are likely to be at odds with
the scenarios for creation of the large--scale structures in the Magellanic System
suggested so far. We investigated this hypothesis for the pure tidal models,
as they were the first ones devised
to explain the evolution of the Magellanic System, and the tidal stripping is intrinsically involved in every model
assuming the gravitational interaction. The parameter space for
the Milky Way (MW)--LMC--SMC interaction was analyzed
by a robust search algorithm (genetic algorithm) combined with a fast restricted N--body model of
the interaction.
Our method extended the known variety of evolutionary scenarios satisfying the observed kinematics
and morphology of the Magellanic large--scale structures.
Nevertheless, assuming the tidal interaction, no satisfactory reproduction of the H\,I data
available for the Magellanic Clouds was achieved with the new proper motions.
We conclude that for the proper motion data by Kallivayalil et al., within their 1--$\sigma$ errors, the dynamical evolution
of the Magellanic System with the currently accepted total mass of the MW cannot be explained in the framework of pure tidal models.
The optimal value for the western component of the LMC proper motion 
was found to be \(\mu_\mathrm{lmc}^\mathrm{W} \gtrsim -1.3$\,mas\,yr$^{-1}\) in case 
of tidal models. It 
corresponds to the reduction of the Kallivayalil et al. value for 
\(\mu_\mathrm{lmc}^\mathrm{W}\) by \(\approx\)40\% in its magnitude.
\end{abstract}


\section{Introduction}\label{intro}
The discussion of the origin and evolution of the Magellanic System has become very intense since
the new proper motion data for the selected LMC/SMC stars were acquired by the Hubble Space Telescope (HST).
The HST measurements yielded the new values for the mean proper motions of the Magellanic Clouds with an unprecedented
accuracy \citep[see][]{KalliLMC, KalliSMC}. In comparison with the previous observational
studies \citep[e.g.][]{Jones94, Kroupa94, Kroupa97}, the corresponding measurement errors were reduced by a factor of 10.
Even though the latest proper motion estimates are consistent with
the previous observational results within the 1--$\sigma$ errors, their actual position in the velocity space
of the Clouds is quite unexpected.

The current velocities of the LMC and the SMC are critical input parameters
of any evolutionary model of the System. However, regarding the large heliocentric distance
to the Magellanic Clouds, nobody attempted the observational determination of their proper
motion until the papers by \cite{Jones94} and \cite{Kroupa94} were carried out.
Some more studies have contributed to the research \citep[e.g.][]{Kroupa97,Drake01,Pedreros02}
offering a span of the mean proper motion values, but reaching no substantial improvement in the
measurement precision, that was still of the same order as the derived values themselves. Such errors
admit a wide variety of scenarios for the interaction \citep{Ruzicka07}, and make
the observational estimates serve only as quite weak tests of the results based
on the theoretical studies of the MW--LMC--SMC interaction \citep[e.g.][]{Lin82, Gardiner94}. 

The theorists focusing on the evolutionary history of the Magellanic System have widely agreed on
two basic physical processes dominating the formation of the Magellanic large--scale structures, including
the Magellanic Stream, the Bridge and the Leading Arm \citep[for details see][]{Bruens05}.
Those are the tidal fields and the ram pressure stripping.

The tidal origin of the extended Magellanic structures was investigated by \cite{Sofue76},
who assumed the LMC and the SMC to form a pair gravitationally bound for several Gyr, moving in
a flattened MW halo. They identified some LMC and SMC orbital
paths leading to the creation of a tidal tail.
\cite{Lin77} pointed out the problem of the large parameter space of the MW--LMC--SMC interaction.
To reduce the size of the parameter space, they neglected both the SMC influence
on the System and dynamical friction within the MW halo, and showed that such
a configuration allows for the existence of a LMC trailing tidal stream.
Following studies by \cite{Murai80}, \cite{Murai84},
\cite{Lin82}, \cite{Gardiner94}, or \cite{Lin95} extended and
developed tidal models of the MW--LMC--SMC interaction.
Generally speaking, the tidal mechanism becomes efficient enough if the timescale
for the interaction is several Gyr \citep{Gardiner94,Gardiner96}.

\cite{Meurer85} involved continuous ram pressure
stripping into their simulation of the Magellanic System. This
approach was followed later by \cite{Sofue94} or by \cite{Moore94}, who simplified
the interaction between the LMC and the SMC, however. The Magellanic Stream was formed of the gas stripped from
outer regions of the Clouds due to collisions with the MW extended
ionized disk. \cite{Heller94} argue for a
LMC--SMC collision resulting into the gas
distribution to the inter--cloud region where it was stripped off by ram
pressure as the Clouds moved through the hot MW halo.
Recently, \cite{Bekki05} applied a
complex gas--dynamical model including star--formation to investigate
the dynamical and chemical evolution of the LMC.
\cite{Mastropietro05} introduced their model of the Magellanic
System including hydrodynamics (SPH) and a full N--body description of
gravity. They studied the interaction between the LMC and the MW.
Neither \cite{Bekki05} nor \cite{Mastropietro05} considered the SMC gas in their models.
However, it was shown that the Stream, which sufficiently reproduces the observed H\,I
column density distribution, might have been created without the SMC gaseous component or even
without the LMC--SMC interaction \citep{Mastropietro05}.
The history of the Leading Arm was not investigated.

In general, hydrodynamical models allow for a better reproduction of the H\,I column density
profile of the Magellanic Stream than tidal schemes. However, they constantly fail
to reproduce the Magellanic Stream radial velocity measurements and especially
the high negative velocity tip of the Magellanic Stream.
Both families of models suffer from serious difficulties when modeling the Leading Arm.
Similar requirements as for the tidal models hold for the
ram pressure stripping schemes concerning interaction timescales,
unless the density of the extended gaseous halo of the MW is increased substantially
over its observational estimates, amplifying the hydrodynamical interaction. The ram pressure force is
proportional to $\upsilon^2$ (relative velocity of the interacting gaseous objects). Thus, the process of gas stripping
becomes more efficient as the velocity increases. However, for finite--size objects, it is accompanied by shorter interaction
timescales due to the reduced crossing time.

Here we came to the actual point of controversy related to the papers by 
\cite{KalliLMC, KalliSMC}:
the HST proper motion values put the Clouds on highly eccentric (or even unbound) orbits around the Galaxy.
\cite{Besla07} analyzed the orbital motion of the LMC using the HST data and brought convincing arguments justifying
the previous statement. Such a result may have serious consequences for the proposed formation mechanisms
of the Magellanic System, since it strongly discriminates the tidal scenario and probably
also the ram pressure--based models. To explain the evolution of the Magellanic
System by either of the mentioned processes, the negative total energy of the Clouds
on their orbits about the MW is needed,
which corresponds to multiple perigalactic approaches over the Hubble time.
It is highly desirable to verify the reliability of the available models of the MW--LMC--SMC interaction
if the orbital angular momentum of the Clouds is as high as found by \cite{KalliLMC, KalliSMC}.

Our paper presents the results of the search of the parameter space for the MW--LMC--SMC interaction
dominated by tides. The approach applied was introduced by \cite{Ruzicka07} and is based on an evolutionary optimization
of the model input according to its ability to reproduce the H\,I observations of the Magellanic System \citep{Bruens05}.
The predictions by \cite{Besla07} and by many others regarding the insufficient performance of tidal models
in case of the \cite{KalliLMC, KalliSMC} proper motions can only be confirmed if the entire parameter space
for the interaction is explored.
The method of the automated search for good models by the genetic algorithms (GA) enabled us to perform such an
analysis for the first time. In addition to the LMC/SMC velocity problem, we also intend to answer two more questions:
Are all the studied parameters of the
same importance for successful modeling the MW--LMC--SMC interaction? Does the evolution of the System show
similar behavior over various scales in the parameter space?

\section{Parameter space of the interaction}
Our model is built in a galactocentric Cartesian frame, assuming the
present position of the Sun \(\mathbf{r}_\odot=(-8.5,0,0)\)\,kpc and its spatial velocity
\(\mathbf{\upsilon}_\odot=(10.0,225.2,7.2)$\,km\,s$^{-1}\) \citep{Dehnen98}.
In total, our study involves over 20 independent
parameters, including the initial conditions of the LMC and the SMC motion, their total masses, parameters of mass distribution,
particle disk radii, and orientation angles, and also the MW dark matter halo flattening parameter.
Some of the parameters were constrained by theoretical studies (including scale radii $\epsilon$ of the LMC/SMC halos,
the Coulomb logarithm $\Lambda$ for the dynamical friction in the MW halo, and the halo flattening
parameter $q$ for the model of the MW gravitational potential). Their mean values
and searched errors were discussed in \cite{Ruzicka07}. This section focuses on the observationally estimated
parameters of the MW--LMC--SMC interaction. 

Figure~\ref{LMC_SMC_pm}
shows the current proper motion of the LMC and the SMC
as estimated by various observational methods, together with
the portion of the velocity space that we studied. 
For convenience, the proper motion vectors were decomposed into
the northern ($\mu^\mathrm{N}$) and the western ($\mu^\mathrm{W}$) components 
\citep[see][]{KalliLMC}.
We included every case that is acceptable up to date.
In case of the LMC proper motion, the measurements by \cite{Jones94, Kroupa94, Kroupa97, Drake01, Pedreros02}
and \cite{KalliLMC} were considered. The proper motion errors for the SMC are based on the results
by \cite{Kroupa97, Irwin99, Freire03, Anderson04a, Anderson04b} and \cite{KalliSMC}.

Efficiency and reliability of the optimization method -- genetic algorithms  -- always depend
on the number of possible solutions, that is extremely high in case of the Magellanic parameter space.
To resolve the difficulty we performed a two--level search. First, the full volume of the parameter space was explored
(referred to as a "global scale" hereafter) to obtain a low resolution information about the behavior
of the tidal model for various combinations of the input parameters.
Subsequently, we reduced the volume of the parameter space by a factor of $10^4$ ("local scale" hereafter)
to study only the 1--$\sigma$ proper
motion errors by \cite{KalliLMC, KalliSMC} with high resolution. Finally, both searches were compared.
The extended ranges of the LMC and the SMC proper motion were
\begin{eqnarray}
\mu^\mathrm{N}_\mathrm{lmc} & = & \langle -0.50,+1.80 \rangle\,\mathrm{mas\,yr^{-1}}
\label{lowres_pm_1}\\
\mu^\mathrm{W}_\mathrm{lmc} & = & \langle -2.23,-0.70 \rangle\,\mathrm{mas\,yr^{-1}}
\label{lowres_pm_2}\\
\mu^\mathrm{N}_\mathrm{smc} & = & \langle -2.50,-0.46 \rangle\,\mathrm{mas\,yr^{-1}}
\label{lowres_pm_3}\\
\mu^\mathrm{W}_\mathrm{smc} & = & \langle -2.05,+0.00 \rangle\,\mathrm{mas\,yr^{-1}}.
\label{lowres_pm_4}
\end{eqnarray}
To perform the high--resolution search, the volume of the parameter space was reduced
by adopting the proper motion values derived by \cite{KalliLMC, KalliSMC}:
\begin{eqnarray}
\mu^\mathrm{N}_\mathrm{lmc} & = & \langle +0.39,+0.49 \rangle\,\mathrm{mas\,yr^{-1}}
\label{hires_pm_1}\\
\mu^\mathrm{W}_\mathrm{lmc} & = & \langle -2.11,-1.95 \rangle\,\mathrm{mas\,yr^{-1}}
\label{hires_pm_2}\\
\mu^\mathrm{N}_\mathrm{smc} & = & \langle -1.35,-0.99 \rangle\,\mathrm{mas\,yr^{-1}}
\label{hires_pm_3}\\
\mu^\mathrm{W}_\mathrm{smc} & = & \langle -1.34,-0.98 \rangle\,\mathrm{mas\,yr^{-1}}.
\label{hires_pm_4}
\end{eqnarray}
The above introduced proper motion space for the Magellanic Clouds is illustrated by Fig.~\ref{LMC_SMC_pm}.

Unlike the proper motion of the Clouds, their LSR radial velocities could be measured with high accuracy.
Following \cite{Vandermarel02}, we set \(\upsilon^\mathrm{rad}_\mathrm{lmc}=262.2\pm 3.4\)\,km\,s\(^{-1}\). The
SMC radial velocity error was estimated by \cite{Harris06} as
$\upsilon^\mathrm{rad}_\mathrm{smc}=146.0\pm 0.6$\,km\,s$^{-1}$.

The heliocentric position vector of the LMC was adopted from \cite{Vandermarel02}, i.e.
the equatorial coordinates are
$(\alpha_\mathrm{lmc}, \delta_\mathrm{lmc})=(81.90^\circ\pm 0.98^\circ, -69.87^\circ\pm 0.41^\circ)$,
its distance modulus is ${(m-M)}_\mathrm{lmc}=18.5\pm 0.1$. 
The equatorial coordinates of the SMC were set to the ranges
$(\alpha_\mathrm{smc}, \delta_\mathrm{smc})=(13.2^\circ\pm 0.3^\circ, -72.5^\circ\pm 0.3^\circ)$
\citep[see][and references therein]{Stanimirovic04}.
\cite{Vandenbergh00} provided a great compilation of various distance determinations for the SMC,
and we used his resulting distance modulus ${(m-M)}_\mathrm{smc}=18.85\pm 0.10$.

Several observational determinations of the LMC disk plane orientation have been published
so far \citep[see, e.g.][]{Lin95}. In our parameter study, the LMC inclination $i$
and position angle $p$ together with their errors agree with \cite{Vandermarel02},
i.e. $i=34.7^\circ\pm 6.2^\circ$ and $p=129.9^\circ\pm 6.0^\circ$. 
As the SMC misses a well defined disk, the orientation and the position angle usually refer to the
SMC "bar" defined by \cite{Gardiner96}. Based on the estimates by \cite{Vandenbergh00} or 
\cite{Stanimirovic04},
we adopted the error ranges $i=60^\circ\pm 20^\circ$ and $p=45^\circ\pm 20^\circ$ for the SMC initial disk inclination
and position angle, respectively.

\cite{Gardiner94} analyzed the H\,I surface contour
map of the Clouds to estimate the initial LMC and SMC disk radii
$r^\mathrm{disk}_\mathrm{lmc}$ and $r^\mathrm{disk}_\mathrm{smc}$, respectively.
Regarding the absence of a clearly defined disk of the SMC and
possible significant mass redistribution in the Clouds during their
evolution, the results require a careful treatment.

Current total masses $m_\mathrm{lmc}$ and $m_\mathrm{smc}$ follow the estimates by 
\cite{Vandenbergh00}.
The masses of the Clouds are functions of time and evolve due to the LMC--SMC exchange of matter,
and as a consequence of
the interaction between the Clouds and the MW. Our test--particle model does not allow for a reasonable treatment
of a time--dependent
mass--loss. Therefore, the masses of the Clouds are considered constant in time, and their initial values at the starting epoch
of simulations are approximated by the current LMC and SMC masses.

The dynamics of the MW--LMC--SMC interaction is critically dependent on the density distribution and the total mass of the Galaxy.
We model the MW by the simple axially symmetric logarithmic potential involving 3 parameters 
\citep{Binney87}.
Only the MW halo flattening $q$ was treated as a free parameter in this study, varying within the range
$\langle0.78,1.20\rangle$ \citep[see also][]{Ruzicka07}, and thus introducing a spread in the total mass of the Galaxy
$m_\mathrm{MW}(q) = \langle 0.92,2.15\rangle\cdot 10^{12}$\,M$_\odot$ within the radius of 200\,kpc.

\section{Methodology}
We investigate the pure tidal models in this paper, as they were the first ones devised to explain the creation of
the Magellanic Stream \citep{Sofue76}. Beside that, the tidal stripping is intrinsically involved in every model
assuming the gravitational interaction.
The model itself is an advanced version of the scheme by \cite{Sofue76}: it is a restricted N--body (i.e. test particle)
code describing the gravitational interaction between the Galaxy and its dwarf companions. The potential of
the MW is dominated by the flattened dark matter halo, and the dynamical friction is exerted on the Magellanic
Clouds as they move through the halo. The LMC and the SMC are represented by Plummer spheres, initially
surrounded by test--particle disks. For further details see \cite{Ruzicka07}.
\subsection{Genetic algorithm search}
As mentioned already, the search itself was performed by GAs that mimic the selection strategy of the natural evolution.
\cite{Holland75} first proposed the application of such an approach on optimization problems in mathematics.
Recently, performance of GAs was studied for galaxies in interaction \citep[see][]{Wahde98,Theis99}.
As an example, \cite{Theis01} analyzed the parameter space
of two observed interacting galaxies -- NGC\,4449 and DDO\,125. GAs turned out to be very robust tools for such a task if the routine comparing
the observational and modeled data is appropriately defined. The approach by 
\cite{Theis01} was later adopted and improved
in order to explore the interaction of the Magellanic Clouds and the Galaxy 
\citep{Ruzicka07}. The comparison between the model and
observations became more efficient by involving an explicit search for the structural shapes in the data. Also
the significant system--specific features (such as a special geometry and kinematics) were taken into account,
further improving the GA performance for exploration of the MW--LMC--SMC interaction. More detailed information is to be found
in Sec.~\ref{understand_F} of this paper.

\subsection{Comparison between model and observations: the Fitness Function}\label{understand_F}
The proposed automatic search of the parameter space is driven by the routine comparing the
modeled and observed H\,I distribution in the Magellanic System \citep{Bruens05}. The match is measured by
the \emph{fitness function} ($F$) which is, in fact, a function of all input parameters, since
every parameter set determines the resulting simulated H\,I data--cube. The devised function $F$
returns a floating--point number between 0.0 (complete disagreement) and 1.0 (perfect match),
and consists of three different comparisons, including search for structures and analysis of local kinematics.

Efficiency of the GA is critically dependent on the applied fitness function.
\cite{Theis01} proposed a generally applicable technique based on comparing the relative intensities of the corresponding
pixels in the modeled and observed data--cubes. Such a fitness function was successfully used to analyze
an interaction involving two galaxies \citep{Theis01}. The mentioned comparison scheme became one of the three components
of the fitness function developed for this project. Both modeled and observed H\,I column
density values are scaled relative to their maxima to introduce
dimensionless quantities. Then, we get
\begin{equation}
F_1 = \frac{1}{N_{\upsilon} \cdot N_x \cdot N_y}\sum\limits_{i=1}^{N_{\upsilon}} \sum\limits_{j=1}^{N_y} \sum\limits_{k=1}^{N_x}
\frac{1}{1 + \left|\sigma_{ijk}^\mathrm{obs} - \sigma_{ijk}^\mathrm{mod}\right|},
\label{fitness1}
\end{equation}
where $\sigma_{ijk}^\mathrm{obs}$, $\sigma_{ijk}^\mathrm{mod}$ are
normalized column densities measured at the position [j, k] of the
i--th velocity channel of the observed and modeled data, respectively.
$N_{\upsilon} = 32$ is the number of separate LSR radial velocity channels in our data.
$(N_x \cdot N_y) = (32 \cdot 64)$ is
the total number of positions on the plane of sky for which observed and modeled H\,I column density values are available.

\cite{Ruzicka07} further tested the performance of GA
for the problem of galactic interactions, and an additional comparison dealing with the
whole data--cubes was devised. It combines the enhancement of structures in the data by their Fourier
filtering with the subsequent check for empty/non--empty pixels in both data--cubes.
The corresponding component of the fitness function is defined as follows:
\begin{equation}
F_2 = \frac{\sum\limits_{i=1}^{N_{\upsilon}} \sum\limits_{j=1}^{N_y} \sum\limits_{k=1}^{N_x} 
\mathrm{pix}_{ijk}^\mathrm{obs} \cdot \mathrm{pix}_{ijk}^\mathrm{mod}}
{\max\left(\sum\limits_{i=1}^{N_{\upsilon}} \sum\limits_{j=1}^{N_y} \sum\limits_{k=1}^{N_x} 
\mathrm{pix}_{ijk}^\mathrm{obs},
\sum\limits_{i=1}^{N_{\upsilon}} \sum\limits_{j=1}^{N_y} \sum\limits_{k=1}^{N_y} 
\mathrm{pix}_{ijk}^\mathrm{mod}\right)},
\label{fitness2}
\end{equation}
where $\mathrm{pix}_{ijk}^\mathrm{obs}\in\{0,1\}$ and $\mathrm{pix}_{ijk}^\mathrm{mod}\in\{0,1\}$ indicate whether there is matter
detected at the position [i, j, k] of the 3\,D data on the observed and modeled Magellanic System, respectively.

Effectively, such a comparison
is a measure for the agreement of the structural shape in the data.
No attention is paid to
specific H\,I column density values here. We only test whether both
modeled and observed emission is present at the same pixel of the
position--velocity space.
\cite{Ruzicka07} showed that the search for structures
significantly improves the GA performance if the structures of interest occupy only a small fraction of the system's
entire data--cube ($<10$\% in the case of the Magellanic Stream and the Leading Arm).

\cite{Ruzicka07} also recommended and
successfully applied a system--specific comparison.
In case of the Magellanic Clouds, the very typical linear radial velocity
profile of the Stream including its high negative velocity tip was considered important.
The slope of the LSR radial velocity
function is a very specific feature, especially strongly dependent on
the features of the orbital motion of the Clouds.
Then, the third $F$ component is defined as
\begin{equation}
F_3 = \frac{1}{1 + \left|\frac{\upsilon_\mathrm{min}^\mathrm{obs} - \upsilon_\mathrm{min}^\mathrm{mod}}{\upsilon_\mathrm{min}^\mathrm{obs}}\right|},
\label{fitness3}
\end{equation}
where $\mathrm{\upsilon_{min}^{obs}}$ and
$\mathrm{\upsilon_{min}^{mod}}$ are the minima of the observed LSR
radial velocity profile of the Magellanic Stream and its model, respectively.
The resulting fitness function $F$ combines the above defined components in the following way:
\begin{equation}
F = F_1 F_2 F_3.
\label{fitnessA}
\end{equation}

In principle, the GA is able to find the global maximum of $F$ (i.e. the best model over the studied
parameter space), but such a process may be very time--consuming due to the possibly slow convergence of
$F$ \citep[see][]{Holland75,Goldberg89}. In order to overcome such a difficulty, we searched the parameter space repeatedly
in a fixed number of optimization steps, i.e. generations of models, and collected 120 high--quality
models for either global or local scale of the parameter space. 
Distribution of the 120 local peaks of $F$ helps to map the fitness function landscape but it
does not allow for conclusions on the behavior of $F$ either outside or inside the volume
populated by the localized models. Therefore, in the following paragraphs we intend to devise a reasonable
method for the further analysis of the fitness function $F$.

At this point, the reader might ask why there is such an attention paid to the properties
of the fitness function itself if it, in fact, does not seem to provide any
physical information about the interacting system of the Galaxy and the Magellanic Clouds.
Indeed, the function $F$ serves
primarily as a driver to the GA engine. However, the search for
good models of the observed Magellanic System is efficient only if relevant astrophysical data
are supplied as the input to $F$. As already mentioned, our study deals with detailed morphological
and kinematic information from the 21\,cm survey by \citet{Bruens05}
and with the corresponding modeled data. The fitness function then makes a link between the observable
data and the initial state of the Magellanic System.
Here we came to the benefits of spending time on studying the function $F$: in principle,
it allows for identification of all points/regions in the parameter space leading to reproduction of
the observational data, and also an appropriate analysis of its behavior may evaluate sensitivity of the System to
variations in different parameters, i.e. their importance to the evolution of the System.

The parameter space of the interacting Magellanic Clouds is very extended.
Acquiring helpful information about the function $F$ is quite a demanding task, but we consider
it feasible once the goals of such an analysis are properly defined.
The investigation of the fitness function $F$ is supposed
to help us to answer two questions already raised in Section~\ref{intro}:
Are all the studied parameters
of the same importance for the properties of $F$? Does $F$ of the system show similar behavior
over various scales in the parameter space?

\subsection{Application of the Fitness Function}
As the first step towards better understanding of the fitness function,
we studied the 1\,D projections of $F$ to the plane of the j--th parameter
\begin{equation}
F^i(p_j) \equiv f(p^i_1,\ldots,p_j,\ldots,p^i_n),
\end{equation}
where ${p^i_1,\ldots,p^i_{j-1},p^i_{j+1},\ldots,p^i_n}$
are the specific values of the parameters corresponding to the \mbox{i--th} GA fit (point in the parameter space)
and the parameter $p_j$ is varied within a given range.
To quantify the sensitivity of $F^i(p_j)$ to changes in different variables (parameters)
several functions were defined. First, we have
\begin{equation}
\Delta^i_j \equiv \frac{\Sigma_j^i}{\overline{F^i(p_j)}},
\label{relative_dev}
\end{equation}
which is the relative deviation of the 1\,D projected fitness $F^i(p_j)$ from its mean value
$\overline{F^i(p_j)}$ on the studied interval, where
\begin{equation}
\Sigma_j^i \equiv \sqrt{\frac{1}{N_\mathrm{P}}\sum\limits_{k=0}^{N_\mathrm{P}-1}(F^i(p_{j,k})-\overline{F^i(p_j)})^2}
\label{standard_dev}
\end{equation}
is the corresponding deviation from $\overline{F^i(p_j)}$ expressed for $N_\mathrm{P}$ points $p_{j,k}$
of the projected function $F^i(p_j)$.
We also calculate the relative change in $F^i(p_j)$ as
\begin{equation}
\Delta^i_{j,\mathrm{F}} \equiv \frac{1}{2}\frac{\max(F^i(p_j)) - \min(F^i(p_j))}{\overline{F^i(p_j)}}.
\label{relative_change}
\end{equation}

The behavior of $F$ is analyzed here in terms of its deviation from the reference levels,
which are established by the mean values of the 1\,D projections $\overline{F^i(p_j)}$.
We found such an approach particularly useful
if one wishes to distinguish between the large--scale and localized significant changes in $F$.

Low values of both, $\Delta^i_j$ and $\Delta^i_{j,\mathrm{F}}$, indicate presence of a global plateau of $F^i(p_j)$,
and thus weak dependence of the system on the j--th parameter.
If the corresponding $\Delta^i_{j,\mathrm{F}}$ is of a significantly higher value, local peaks (or wells) exist.
Similarly, the overall considerable evolution of $F^i(p_j)$
is revealed by high values of both $\Delta^i_j$ and $\Delta^i_{j,\mathrm{F}}$. In such a case
additional information is provided by the ratio of the functions (\ref{relative_dev}) and (\ref{relative_change}).
As the ratio approaches unity, abrupt changes in $F^i(p_j)$ are favored over its smooth and slow evolution.

The intention of this section is to apply the method for the fitness function analysis introduced
in Sec.~\ref{understand_F} to the Magellanic System. The functions $\Delta^i_j$ and $\Delta^i_{j,\mathrm{F}}$
are helpful if features and behavior of the fitness function $F$ are studied in the neighborhood of an arbitrary
point in the parameter space. However, to answer the above raised questions, an approach somewhat less detailed
is sufficient. We suggest to treat the functions $\Delta^i_j$ and $\Delta^i_{j,\mathrm{F}}$ statistically and
to calculate their mean values $\Delta_j \equiv \overline{\Delta^i_j}$ and
$\Delta_{j,\mathrm{F}} \equiv \overline{\Delta^i_{j,\mathrm{F}}}$ over all 120 GA fits for both
global and local scale cases. The function $F$ may be characterized by significantly different
values of $\Delta^i_j$ or $\Delta^i_{j,\mathrm{F}}$ depending on the selected point in the parameter space.
But for now, we are particularly interested in general trends in the behavior of the fitness function (i.e. the
behavior of our model for the interaction) that should be expected if one studies the impact of
variations in a selected parameter. As we will learn later, the identified
GA fits cover a large fraction of the total parameter space volume. This fact also justifies the proposed statistical
treatment of the 1\,D projections of the function $F$ in case the results apply on the entire parameter space.

\section{Results}\label{results}
Regarding the immense difficulties accompanying the observational measurement of the proper motions
in case of the Magellanic Clouds, the models of the MW--LMC--SMC interaction were used to draw conclusions
on the motion of the Clouds \citep[e.g.][]{Sofue76,Gardiner94,Heller94, Gardiner96}.
The mentioned models preferred either the tidal or hydrodynamical interactions as
the processes dominating the evolution of the Magellanic System.
Generally speaking, the proposed formation mechanisms
are not efficient enough unless the Clouds orbit around the Galaxy.
Several perigalactic approaches of the Clouds are expected
by the tidal models \citep{Gardiner94, Connors05}. Shorter timescales for the interaction
may be sufficient within the ram pressure scenario \citep{Heller94,Mastropietro05}.
However, the proper motions by \cite{KalliLMC,KalliSMC} lead to timescales further dramatically reduced,
as the Clouds should be approaching the Galaxy for the first time \citep{Besla07}.
The research of the dynamical evolution of the Magellanic System is at the point
where our theoretical understanding of the MW--LMC--SMC interaction is at odds
with some critical observational constraints. 

\cite{Ruzicka07} realized that the previous attempts to model the Magellanic System
always reduced the parameter and initial condition space for the interaction
by additional assumptions, such as omitting the SMC, or adopting a special orbit
for the LMC. However, the uniqueness of the models is unclear unless a systematic
analysis of the entire parameter space compatible with the available observations
is performed. The idea by \cite{Ruzicka07} was justified as they used
a tidal model of the interaction and reproduced the observed H\,I structures
for remarkably different histories of the System. Such a complex approach
requires a powerful search method.
\cite{Ruzicka07} resolved the difficulty by employing GAs as optimization tools characterized
by reliability and low sensitivity to local extremes \citep{Theis01}. We adopted
their approach to analyze the performance of pure tidal models in case of the Magellanic System
assuming the LMC and the SMC proper motions by \cite{KalliLMC,KalliSMC}.

\subsection{Parameter Dependence}
The values of $\Delta_j$ and $\Delta_{j,\mathrm{F}}$ were calculated for the fitness function $F$
over both global and local scales of the parameter space. Table~\ref{table_F}
presents the result in the descending order according to the value of the function $\Delta_j$.
Following the previous explanation it is clear, that the order reveals the sensitivity of the fitness
function to various parameters. At this point, we feel qualified to answer the first question raised in
Sec.~\ref{understand_F}: the parameters are not of the same importance for the properties of $F$.
Table~\ref{table_F} indicates that the sensitivity of the Magellanic System to the choice for the LMC/SMC proper motions
is significantly higher than in case of any of the remaining parameters
This conclusion is
well illustrated by Figure~\ref{2dvelocityLMC} depicting the fitness landscape projected
to the plane of the LMC proper motion components. The projection was made at the positions of the
best GA fits found for the global and local scale cases, respectively.

Our analysis has shown the critical dependence of
the evolution of the Magellanic System on the LMC and the SMC spatial velocities.
The function \(\Delta_{j,\mathrm{F}}\) is not only useful for the detailed
analysis of the fitness function, but can also serve to estimate the typical overall change in the fitness value
over the entire studied parameter range. The value of \(\Delta_{j,\mathrm{F}}\) exceeds 0.15 for every
LMC/SMC proper motion component on the global scale (Tab.~\ref{table_F}).
Regarding the definition by Eq.~(\ref{relative_change}),
the mentioned values multiplied by the factor of 2.0 tell us how much
the 1\,D fitness projections $F^i(p_j)$ change typically compared to their mean values. As the usual mean value
$\overline{F^i(p_j)}$ equals to $\approx 0.3$, one may assume the global proper motion ranges
given by Eq.~(\ref{lowres_pm_1}) to~(\ref{lowres_pm_4}) to derive the following estimated changes in the
fitness of our models per the unit step in the proper motion value:
\begin{eqnarray}
\Delta F(\mu^\mathrm{N}_\mathrm{lmc})/\Delta\mu^\mathrm{N}_\mathrm{lmc} & \simeq & 0.04\,\mathrm{mas^{-1}\,yr}
\label{fitness_step_1}\\
\Delta F(\mu^\mathrm{W}_\mathrm{lmc})/\Delta\mu^\mathrm{W}_\mathrm{lmc} & \simeq & 0.06\,\mathrm{mas^{-1}\,yr}
\label{fitness_step_2}\\
\Delta F(\mu^\mathrm{N}_\mathrm{smc})/\Delta\mu^\mathrm{N}_\mathrm{smc} & \simeq & 0.09\,\mathrm{mas^{-1}\,yr}
\label{fitness_step_3}\\
\Delta F(\mu^\mathrm{W}_\mathrm{smc})/\Delta\mu^\mathrm{W}_\mathrm{smc} & \simeq & 0.06\,\mathrm{mas^{-1}\,yr}
\label{fitness_step_4}
\end{eqnarray}
The above listed values should be treated very carefully, as they simplify the real behavior of the fitness function.
The fact that the proper motion of the SMC is even more critical than that of the LMC indicates, that both Clouds
serve as sources of matter for the Magellanic large scale structures, but the SMC contribution responds to the choice
for the orbit more strongly.
It is another nice illustration of how much information about the physical properties of the Magellanic System
can be actually obtained from the fitness function.

The estimates given by the Eq.~(\ref{fitness_step_1}) to~(\ref{fitness_step_4}) give a hint at
the relation between the 
fitness and the physical properties of our models. Later in Sec.~\ref{spatial_motion} the lower limit
for the fitness of the satisfactory models will be established and discussed. While the best model we found
yields $F_\mathrm{best}=0.514$, the fitness threshold level is $F_\mathrm{lim}>0.434$, i.e. all the acceptable
models are found within the fitness range $(F_\mathrm{best}-F_\mathrm{lim})=0.08$. If any of the LMC/SMC proper motion components
changes by 1\,mas\,yr$^{-1}$, the corresponding model is extremely likely to cross the border between the successful and
insufficient models.
If some of the remaining
parameters are altered by further observations, our findings concerning the motion of the Clouds
shall not be affected strongly.

\subsection{Scale Dependence}
What if the volume of the parameter space is reduced substantially by changing the proper motion ranges
of the Clouds? To find the answer, let's take a look at Table~\ref{table_F}. It indicates that the most influential
parameters remain the same, if we zoom into the velocity space. The role of the choice for the LMC/SMC proper motion
is still dominant for the evolution of the System. Impact of different proper motion components may be altered if
their searched ranges are reduced.
It is also a notable fact that the values of both functions $\Delta_j$ and $\Delta_{j,\mathrm{F}}$
decreased systematically over all studied parameters as we switched to the local scale of the parameter space.
It means that the fitness function $F$ does not have a fractal--like (irregular) structure,
and the probability of missing steep
high peaks (i.e. isolated quality models) in the fitness landscape may be reduced by running the GA search again
on that sub--region of the original parameter space, which is of a particular interest.
We demonstrated that the effective resolution of the GA search
may be improved by reducing the volume of the parameter space. Therefore, the exploration of the parameter
space for the MW--LMC--SMC interaction was performed in two levels. First, every LMC/SMC proper motion
estimate available was included. Subsequently, the proper motion spread was reduced to the 1--$\sigma$ error ranges
by \cite{KalliLMC, KalliSMC} to verify or eventually correct the global scale search.

\subsection{Spatial motion and tidal models}\label{spatial_motion}
We have analyzed the influence of the involved parameters on the tidal interaction in the Magellanic System.
Reproduction of the H\,I observational data \citep{Bruens05} by the restricted N--body simulation
turned out to be critically dependent on the current spatial velocities of the Clouds.

Since the effective resolution of the search can be improved by reducing the volume of the studied
parameter space (as justified earlier in this section), two sets of the LMC/SMC proper motion 
ranges were assumed. After including every proper motion estimate available up to date,
only the values by \cite{KalliLMC,KalliSMC} were involved to allow for a high--resolution search on the local scale
of the parameter space.
With the use of GA, $\approx 2\cdot 10^6$ parameter combinations, i.e. individual N--body simulations,
were tested in total, and 120 sets providing
the highest fitness
were collected for each of the studied
volumes of the parameter space.

The resulting models are not considered acceptable unless they produce both
leading and trailing H\,I Magellanic structures (the Leading Arm and the Magellanic Stream).
Thus, to allow for quantitative statements, a threshold level of $F$ had to be established.
In order to do so, one needs to understand how the modeled H\,I morphology and kinematics
is reflected in the value of the fitness function.

The modeled distribution and kinematics of H\,I in the Leading Arm region and around the
main LMC and SMC bodies remains similarly unsatisfactory over the entire parameter space,
especially failing to reproduce the observed
morphology of the Leading Arm (see Fig.~\ref{3d_maps}). In terms of the fitness function, the value of $F$ never exceeds 0.3 if calculated
only for the Leading Arm.

It is the Magellanic Stream that turned to be very sensitive to the choice for the
model parameters, and critically influencing the resulting fitness. Figure~\ref{3d_maps}
illustrates the above discussed facts. While the model of $F=0.514$ (global scale) is able to fit the basic 
features of the Magellanic Stream both in the projected H\,I distribution and the LSR radial velocity profile,
the best model for the increased LMC/SMC spatial velocities ($F=0.336$, local scale) places the Magellanic Stream to the
position--position--velocity space incorrectly, as the $\approx-400$\,km\,s\(^{-1}\) tip of the Stream
is shifted towards the Clouds by \(\approx20^\circ\) compared to the observations
(1 pixel in Fig.~\ref{3d_maps} equals roughly to \(2^\circ\) in the plane of sky),
and the modeled morphology differs seriously from the observed structure of the Magellanic Stream. 
Generally speaking, the described behavior of the modeled trailing stream is responsible
for the resulting fitness of a given model, and was used to define the desired threshold level
of the \(F\) value.

Unlike the local scale analysis, the global search of the entire parameter space
always resulted in a model placing the high negative radial velocity tip of the Stream to the
correct projected position, and the modeled trailing stream filled roughly the same area of the H\,I data--cube
as the observed Magellanic Stream. Following the mentioned fact, we selected the worst of the 120 fits
from the global scale search to represent the threshold level of the fitness and so $F>0.434$ defines
satisfactory models of the Magellanic System.

\begin{figure}
\includegraphics[bb=0 0 550 800,angle=90,scale=.28]{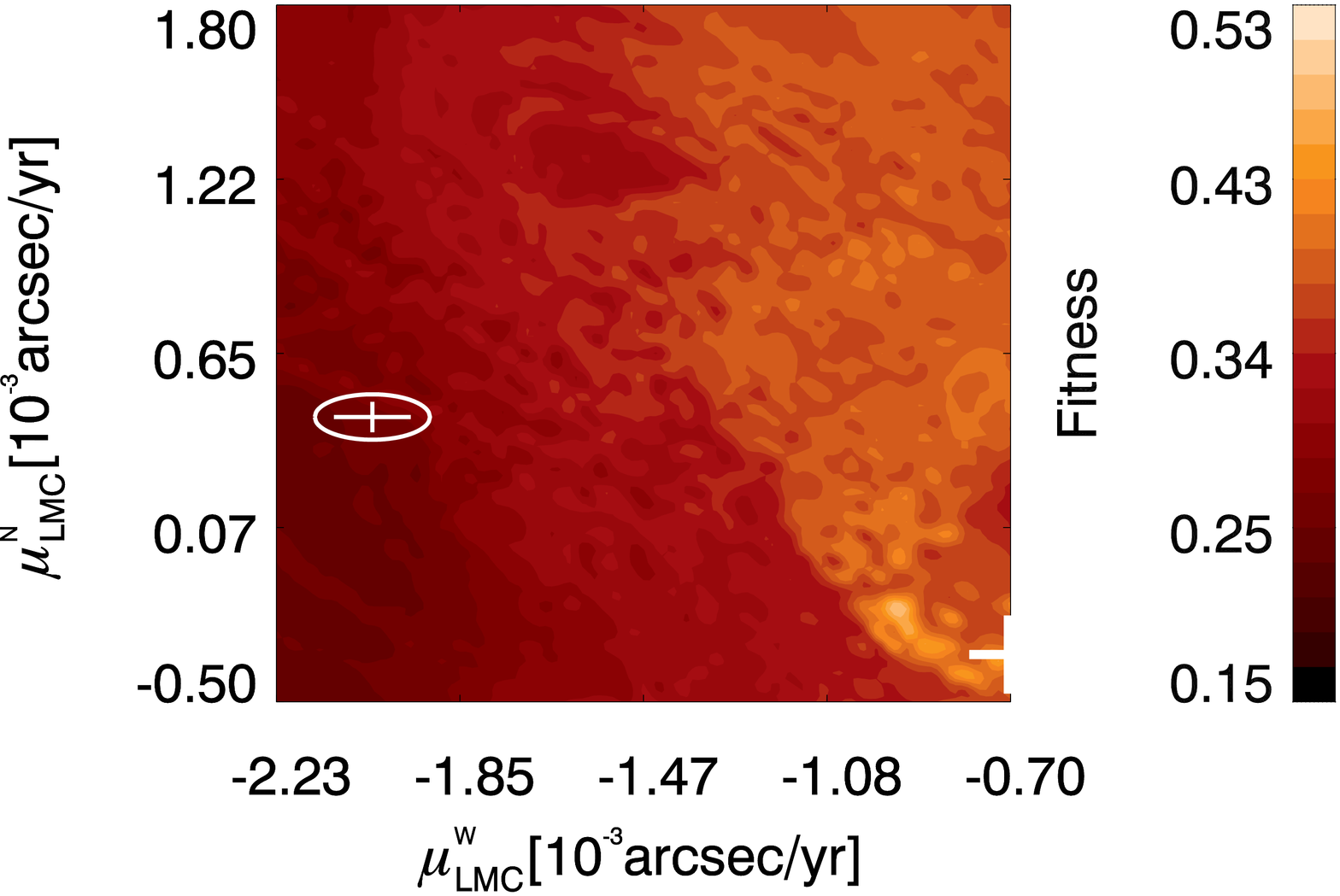}\\
\includegraphics[bb=0 0 550 800,angle=90,scale=.28]{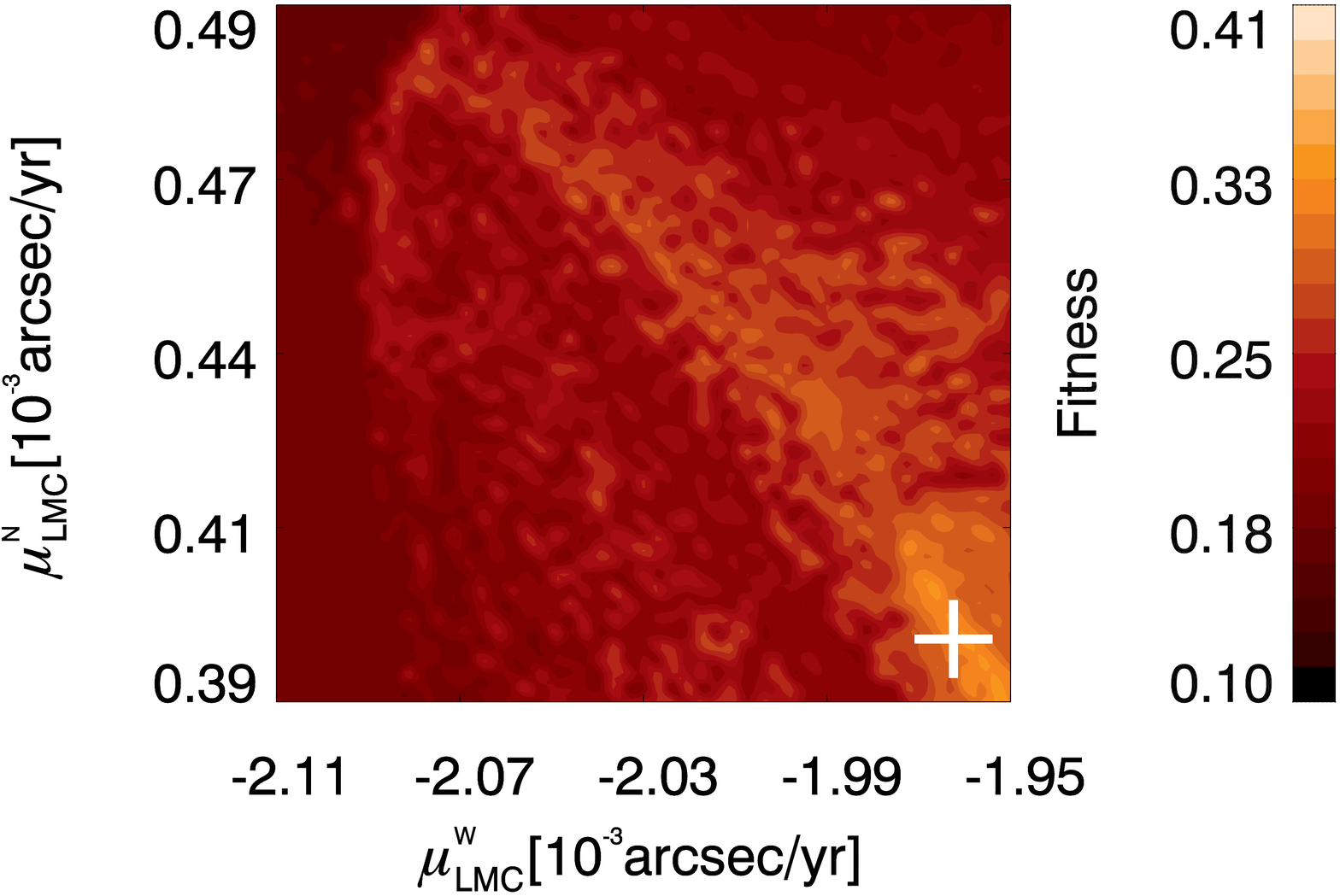}
\caption{The (\(\mu^\mathrm{N}_\mathrm{lmc}\), \(\mu^\mathrm{W}_\mathrm{lmc}\))--planes for the 2\,D
projections of the function \(F\). For the moment, the remaining parameters are fixed to the values
corresponding to the best models for the original (upper plot) or the reduced
proper motion ranges, respectively. The best models are marked by the white crosses.
The white ellipse in the upper plot indicates the 68.3\,\% confidence region of the LMC proper motion by 
\cite{KalliLMC}
and roughly corresponds to the entire (\(\mu^\mathrm{N}_\mathrm{lmc}\), \(\mu^\mathrm{W}_\mathrm{lmc}\))--plane
depicted in the lower plot. 
\label{2dvelocityLMC}}
\end{figure}

Figure~\ref{2dvelocityLMC} indicates, that the
sub--region of the ($\mu_\mathrm{lmc}^\mathrm{N}$, $\mu_\mathrm{lmc}^\mathrm{W}$)--plane
introduced by \cite{KalliLMC} does not allow for any satisfactory models assuming the pure tidal interaction.
Such a conclusion is also confirmed by the local scale search restricted to the HST proper motion data only
(lower plot of Figure~\ref{2dvelocityLMC}).
Thus, Fig.~\ref{2dvelocityLMC} denotes that it is not
possible to simulate the evolution of the Magellanic System by pure tidal models if the spatial
velocity of the LMC is as high as predicted by \cite{KalliLMC}. However, such a speculation can only be
confirmed if the entire parameter space of the interaction is explored, as it is based on the behavior
of the fitness function $F$ in the neighborhood of the selected point.

Figure~\ref{fit-muLMC} summarizes our results. The value of every local peak of the fitness function $F$
(i.e. the fitness of every model identified by GA)
for the interaction of the MW--LMC--SMC system is plotted as function of the
$\mu_\mathrm{lmc}^\mathrm{N}$ and $\mu_\mathrm{lmc}^\mathrm{W}$ proper motion components for the LMC.
While the northern component $\mu_\mathrm{lmc}^\mathrm{N}$ allows for satisfactory
modeling of the Magellanic System over the entire global range, the situation is quite
different for the LMC proper motion in the western direction ($\mu_\mathrm{lmc}^\mathrm{W}$).
The upper right plot of Figure~\ref{fit-muLMC} clearly indicates that the tidal scheme
fails unless the $\mu_\mathrm{lmc}^\mathrm{W}$ values by \cite{KalliLMC}
are reduced by $\approx 40$\,\% to reach $\mu_\mathrm{lmc}^\mathrm{W} \gtrsim -1.3$\,mas\,yr$^{-1}$.
\cite{Besla07} have shown that the proper motion
component $\mu_\mathrm{lmc}^\mathrm{W}$ controls the Galactocentric spatial velocity of the LMC.
Hence, one can easily see that the pure tidal models of the Magellanic System
constantly fail for the spatial velocities putting the LMC on highly eccentric
or maybe unbound orbits about the Galaxy \citep{Besla07}. This conclusion
was verified and confirmed by the local scale parameter study focusing
on the proper motion ranges by \cite{KalliLMC, KalliSMC}. Despite our efforts, no even decent
models were found over this portion of the parameter space (see the lower row of Figure~\ref{fit-muLMC}).
Please note also that the concentration of the GA fits towards lower values of
$\mu_\mathrm{lmc}^\mathrm{W}$ still remains even for the reduced velocity ranges.

Within the original volume of the parameter space, assuming the tidal interaction,
no satisfactory reproduction of the H\,I data
by \cite{Bruens05} was achieved for the HST proper motions.
In agreement with the previous studies, the model succeeded only if the Clouds were moving at
substantially lower Galactocentric velocities. The following high--resolution analysis
of the local scale of the parameter space (the proper motion ranges for the Clouds were restricted to the HST velocity data)
did not change the previous result and no specific quality parameter combinations
were revealed. Regarding the above summarized facts and results, we conclude that the dynamical evolution
of the Magellanic System with the currently accepted total mass of the MW
cannot be explained in the framework of pure tidal models for
the proper motion data by \cite{KalliLMC,KalliSMC} within their 1--$\sigma$ errors.

\section{Conclusions}
The new results introduced by \cite{KalliLMC,KalliSMC} have serious consequences
for our understanding of the dynamical evolution of the Magellanic System, and in 
a wider context
they influence our view of the Local Group and its formation. Such facts, together with the key result of this paper
indicating a conflict between the tidal models and observations of the Magellanic System,
lead us necessarily to the questions about the reliability of the original HST data and
correctness of their treatment by \cite{KalliLMC,KalliSMC}.

Unfortunately, the first issue cannot be addressed until the HST measurements are cross--checked
by an instrument of a competitive resolution and other significant physical characteristics.
In a close future, the GAIA mission should allow for high--precision astrometry, and hopefully
the Magellanic Clouds and their proper motions will become objects of its interest as soon as possible.

Recently, an interesting constraint on the current velocity of the LMC was introduced by 
\cite{Mcclure08}.
They were able to estimate the Galactocentric distance to the cross--section of the Magellanic Leading Arm
with the gaseous disk of the MW. Although the observed part of the Leading Arm does not necessarily
trace the future orbit of the LMC (the Leading Arm is believed to lead the Magellanic System),
the measured position puts the lower limit on the Galactocentric distance
to the point of the next passage of the LMC through the Galactic plane. Following 
\cite{Mcclure08},
the estimated distance is not at odds with the value obtained by adopting the current LMC velocity by 
\cite{KalliLMC}.
In addition to the above discussed issue, a verification of the distances to the main bodies
of the Clouds may be worth a consideration.
Space velocities also depend on the distances to the LMC/SMC, which should be
checked carefully in future to clarify how far they may influence our conclusions.

Concerning the HST proper motion data processing by \cite{KalliLMC, KalliSMC}, both papers indicate that
the data were analyzed and interpreted very carefully and the adopted method is well justified.
A very strong support to the conclusions by \cite{KalliLMC, KalliSMC} came recently from the study
by \cite{Piatek07} who also derived the LMC and the SMC current proper motions from the original HST data.
\cite{Piatek07} introduced a different method to process the LMC/SMC stellar proper motions,
but their results agree with the findings by \cite{KalliLMC, KalliSMC} very well. 
They were able to reduce the LMC/SMC proper motion errors by the factor 
of 3 compared to \cite{KalliLMC, KalliSMC} due to their modified treatment 
of the original data. 
However, while their 1--$\sigma$ proper motion errors of the LMC are completely 
embedded within the corresponding error estimates by \cite{KalliLMC}, 
the mean value of SMC proper motion component $\mu_\mathrm{smc}^\mathrm{W}$ 
is offset by +0.41\,mas\,yr\(^{-1}\) compared to \cite{KalliSMC} (see 
Fig.~\ref{LMC_SMC_pm}).

It means that the region of the parameter space 
delimited by the 1--$\sigma$ proper motion errors by \cite{Piatek07} 
entered our global scale analysis, but it was not explored 
by the local scale search. Nevertheless, there is no indication that 
such a high--resolution exploration might alter significantly the 
results obtained for the global scale search. 
Since the only difference between the results by \cite{KalliLMC, 
KalliSMC} and those by \cite{Piatek07} exists for the western proper 
motion component of the SMC, the Eq.~(\ref{fitness_step_4}) yields a 
typical growth in the models' fitness of $\Delta F(\mu^\mathrm{W}_\mathrm{smc})\approx 
+0.03$ as one switches from \cite{KalliLMC, KalliSMC} 
to \cite{Piatek07}, because the $\mu_\mathrm{smc}^\mathrm{W}$ estimate by 
\cite{Piatek07} 
is by $\approx 0.4$\,mas\,yr\(^{-1}\) lower than the result by \cite{KalliSMC}.
The overall quality of the 
models for the \cite{Piatek07} data is then well below the fitness threshold 
level established in Sec.~\ref{spatial_motion}. Under such conditions it is extremely unlikely to expect any 
significant revision of the global scale parameter search by the high resolution 
analysis of the proper motions by \cite{Piatek07}.

If we admit that the latest observational estimates of the current proper motion for
both Magellanic Clouds are correct, what would be the impact on our understanding
of the Magellanic System and its evolution? We showed that the pure tidal stripping
is insufficient to redistribute the H\,I gas in the System to conform the
available observations \citep{Bruens05}. Nevertheless, the classical model scheme by 
\cite{Sofue76}
offers an extremely simplified view of the physical processes influencing galactic interactions.
Our results, together with the conclusions by \cite{Besla07}, do not allow for more than little
doubts, that despite the high spatial velocities of the Clouds, the models of the interaction
may succeed if sufficiently efficient physical mechanisms are introduced.

In the first order, the ram pressure stripping \citep{Mastropietro05} scenario is well worth further efforts,
since its efficiency strongly depends on several only weakly constrained parameters, namely
the density of the extended gaseous halo of the Galaxy. Moreover, the ram pressure force scales
as $\upsilon^2$ with the relative velocity of the interacting objects, and so the high spatial velocities
of the Clouds by \cite{KalliLMC,KalliSMC} may compensate the effect of the corresponding reduced timescale
for the interaction with the gaseous halo of the Galaxy.
Recently, the paper by \cite{Nidever07}
offered an exciting alternative to the tidal/ram pressure models, considering the
possible intense ejection of mass from several star--forming regions in the LMC super--shells.

{
\acknowledgments
The authors gratefully acknowledge the support
by the Academy of Sciences of the Czech Republic
(the Junior research grant KJB300030801 and the Institutional Research Plan AVOZ10030501),
by the project M\v{S}MT LC06014 Center for Theoretical Astrophysics, and also by the grant P20593--N16
of the FWF Austrian Science Fund.
}

\clearpage

\begin{deluxetable}{lrrclrr}
\tablewidth{0pc}
\tablecolumns{7}
\tabletypesize{\scriptsize}
\tablecaption{Parameter dependence of the fitness function}
\tablehead{
\multicolumn{3}{c}{Global scale} & \colhead{} & \multicolumn{3}{c}{Local scale} \\
\cline{1-3} \cline{5-7} \\
\colhead{\(j\)} & \colhead{\(\Delta_j\cdot 10^{2}\)} & \colhead{\(\Delta_{j,\mathrm{F}}\cdot 10^{2}\)} & \colhead{} &
\colhead{\(j\)} & \colhead{\(\Delta_j\cdot 10^{2}\)} & \colhead{\(\Delta_{j,\mathrm{F}}\cdot 10^{2}\)} \\
}
\startdata
$\mu^\mathrm{N}_\mathrm{smc}$         	& 15.96 & 31.03 & & $\mu^\mathrm{N}_\mathrm{lmc}$		& 3.09 & 5.53 \\
$\mu^\mathrm{W}_\mathrm{lmc}$         	&  8.76 & 17.20 & & $\mu^\mathrm{N}_\mathrm{smc}$		& 3.01 & 5.48 \\
$\mu^\mathrm{W}_\mathrm{smc}$         	&  8.55 & 18.92 & & $\mu^\mathrm{W}_\mathrm{lmc}$		& 2.64 & 4.85 \\
$\mu^\mathrm{N}_\mathrm{lmc}$         	&  6.98 & 15.24 & & $\mu^\mathrm{W}_\mathrm{smc}$		& 2.63 & 4.76 \\
$\alpha_\mathrm{lmc}$ 	              	&  4.22 &  8.34 & & $\alpha_\mathrm{lmc}$			& 1.77 & 3.31 \\
$(m-M)_\mathrm{smc}$		      	&  3.95 &  8.02 & & $R^\mathrm{disk}_\mathrm{lmc}$		& 1.69 & 2.94 \\
$(m-M)_\mathrm{lmc}$ 		     	&  3.92 &  7.91 & & $M_\mathrm{lmc}$				& 1.69 & 3.11 \\
$M_\mathrm{lmc}$ 	              	&  3.86 &  7.78 & & $\epsilon_\mathrm{lmc}$			& 1.67 & 3.09 \\
$M_\mathrm{smc}$ 	              	&  3.48 &  6.93 & & $(m-M)_\mathrm{lmc}$			& 1.65 & 3.06 \\
$\Lambda$ 	                      	&  3.46 &  6.95 & & $\Lambda$					& 1.63 & 2.98 \\
$\upsilon^\mathrm{rad}_\mathrm{lmc}$	&  3.31 &  6.52 & & $(m-M)_\mathrm{smc}$			& 1.59 & 2.95 \\
$\epsilon_\mathrm{lmc}$	  	 	&  2.60 &  5.36 & & $M_\mathrm{smc}$				& 1.49 & 2.76 \\
$\delta_\mathrm{lmc}$		    	&  2.38 &  4.69 & & $R^\mathrm{disk}_\mathrm{smc}$		& 1.47 & 2.64 \\
$\epsilon_\mathrm{smc}$   		&  2.23 &  4.62 & & $\upsilon^\mathrm{rad}_\mathrm{lmc}$ 	& 1.41 & 2.64 \\
$\upsilon^\mathrm{rad}_\mathrm{smc}$	&  1.92 &  3.85 & & $\epsilon_\mathrm{smc}$			& 1.40 & 2.69 \\
$\alpha_\mathrm{smc}$ 	              	&  1.91 &  3.89 & & $\delta_\mathrm{lmc}$            		& 1.03 & 1.90 \\
$i_\mathrm{smc}$			&  1.89 &  3.80 & & $p_\mathrm{smc}$				& 0.99 & 1.80 \\
$\delta_\mathrm{smc}$		    	&  1.83 &  3.73 & & $i_\mathrm{smc}$				& 0.88 & 1.59 \\
$p_\mathrm{smc}$			&  1.75 &  3.62 & & $\alpha_\mathrm{smc}$			& 0.88 & 1.59 \\
$R^\mathrm{disk}_\mathrm{lmc}$		&  1.70 &  3.60 & & $\delta_\mathrm{smc}$            		& 0.86 & 1.54 \\
$R^\mathrm{disk}_\mathrm{smc}$		&  1.55 &  3.30 & & $\upsilon^\mathrm{rad}_\mathrm{smc}$ 	& 0.78 & 1.47 \\
$i_\mathrm{lmc}$ 	         	&  0.70 &  1.39 & & $i_\mathrm{lmc}$				& 0.23 & 0.39 \\
$p_\mathrm{lmc}$	         	&  0.64 &  1.29 & & $p_\mathrm{lmc}$				& 0.21 & 0.37 \\
\enddata
\label{table_F}
\end{deluxetable}
\clearpage

{
\onecolumn
\begin{figure}
\includegraphics[angle=90, scale=.55, clip]{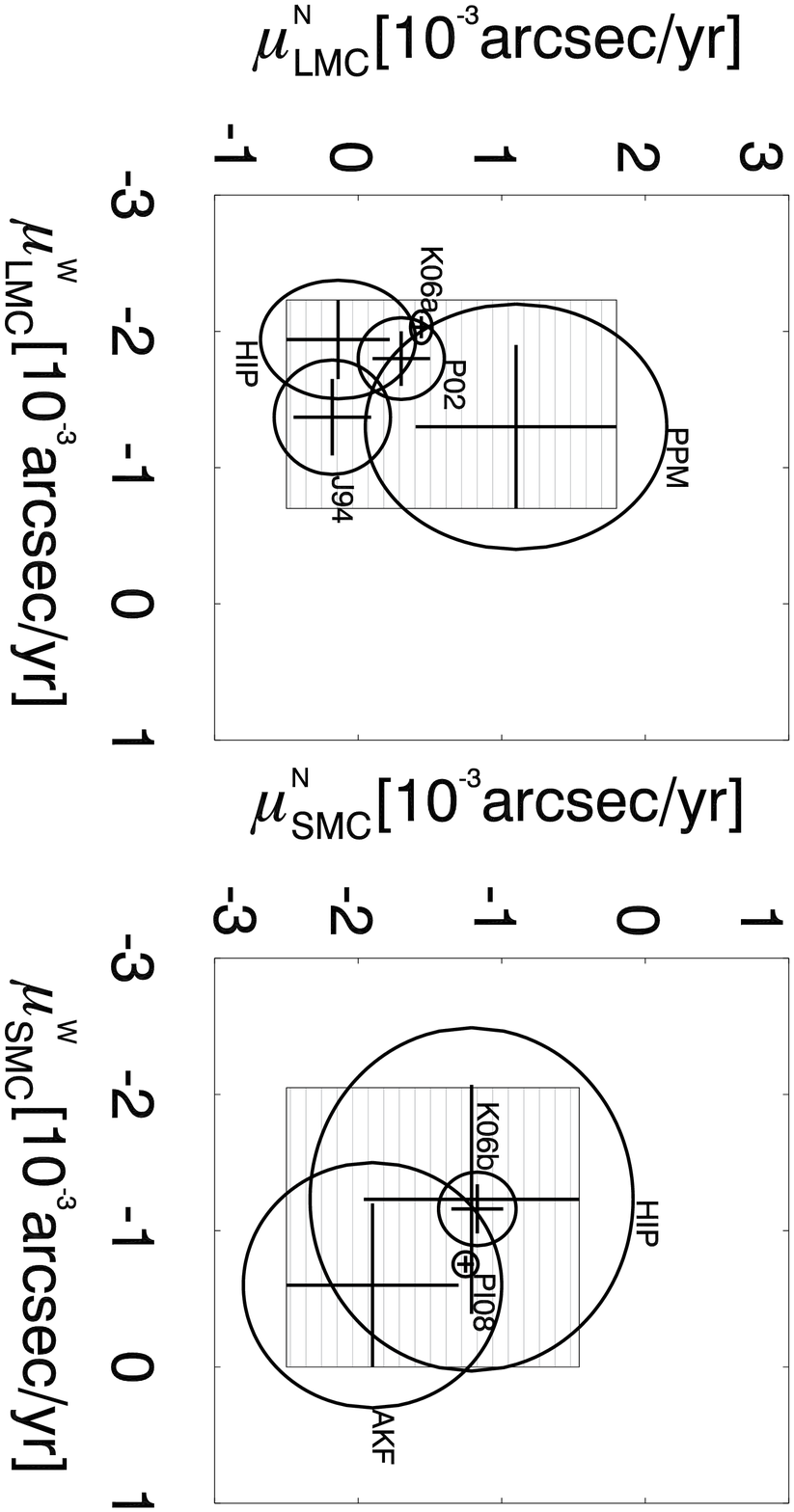}
\caption{The 2\,D projections of the Magellanic parameter space
to the ($\mu^\mathrm{N}$, $\mu^\mathrm{W}$)--plane for both the LMC (left plot) and the SMC.
The gray fillings mark the proper motion ranges explored by the GAs. The labels
indicate the proper motions as expected by various studies.
K06a stands for \cite{KalliLMC}, K06b for \cite{KalliSMC}, PI08 
for \cite{Piatek07}, J94 for \cite{Jones94},
PPM for \cite{Kroupa94}, HIP for \cite{Kroupa97},
P02 for \cite{Pedreros02}, AKF for the value
combining \cite{Freire03} with \cite{Anderson04a, Anderson04b}.
The ellipses show the 68.3\,\% confidence regions.
\label{LMC_SMC_pm}}
\end{figure}
}
\clearpage

{
\onecolumn
\begin{figure}
\includegraphics[bb=30 135 475 735, angle=90, scale=.257, clip]{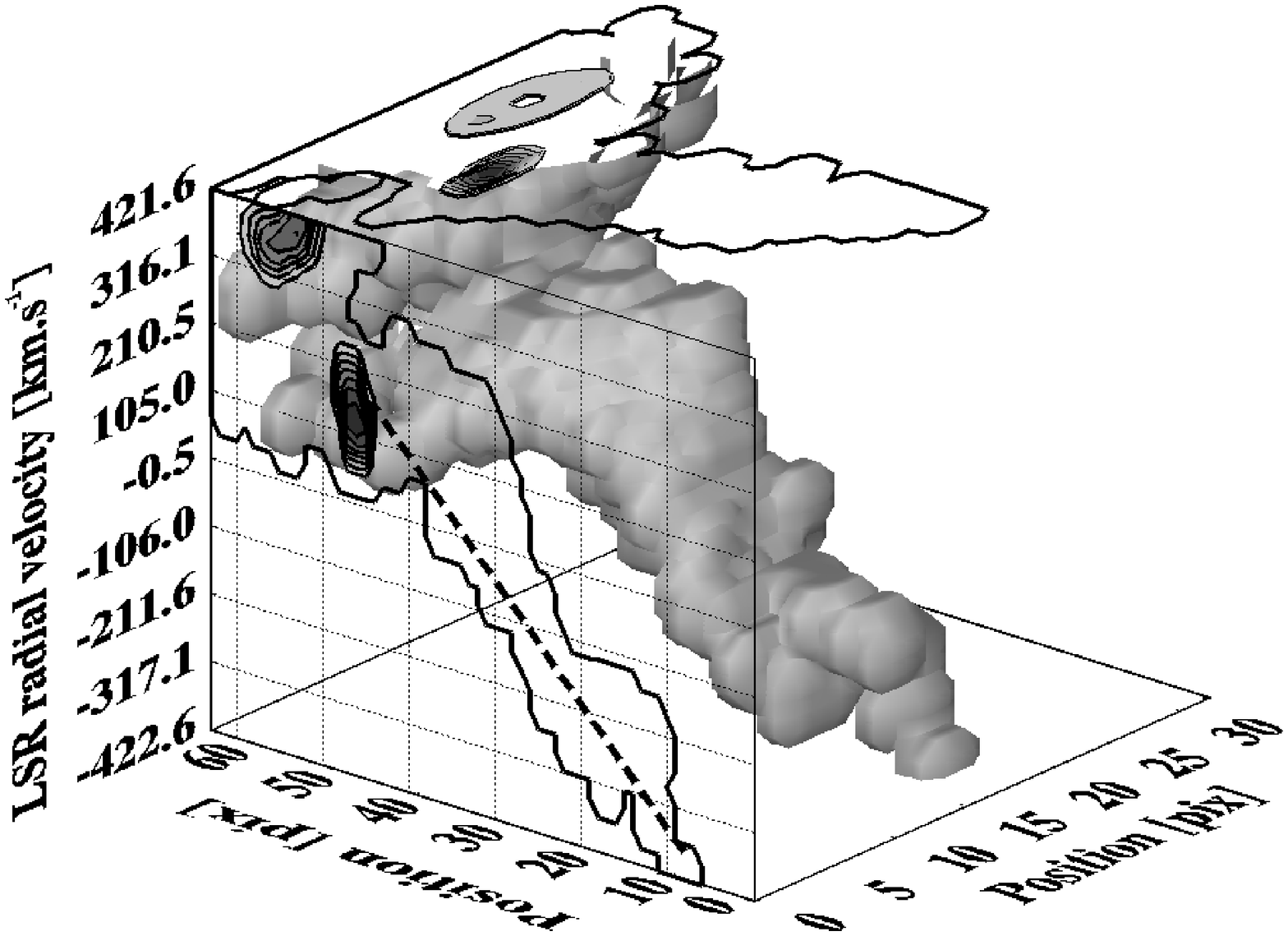}
\includegraphics[bb=30 135 475 735, angle=90, scale=.257, clip]{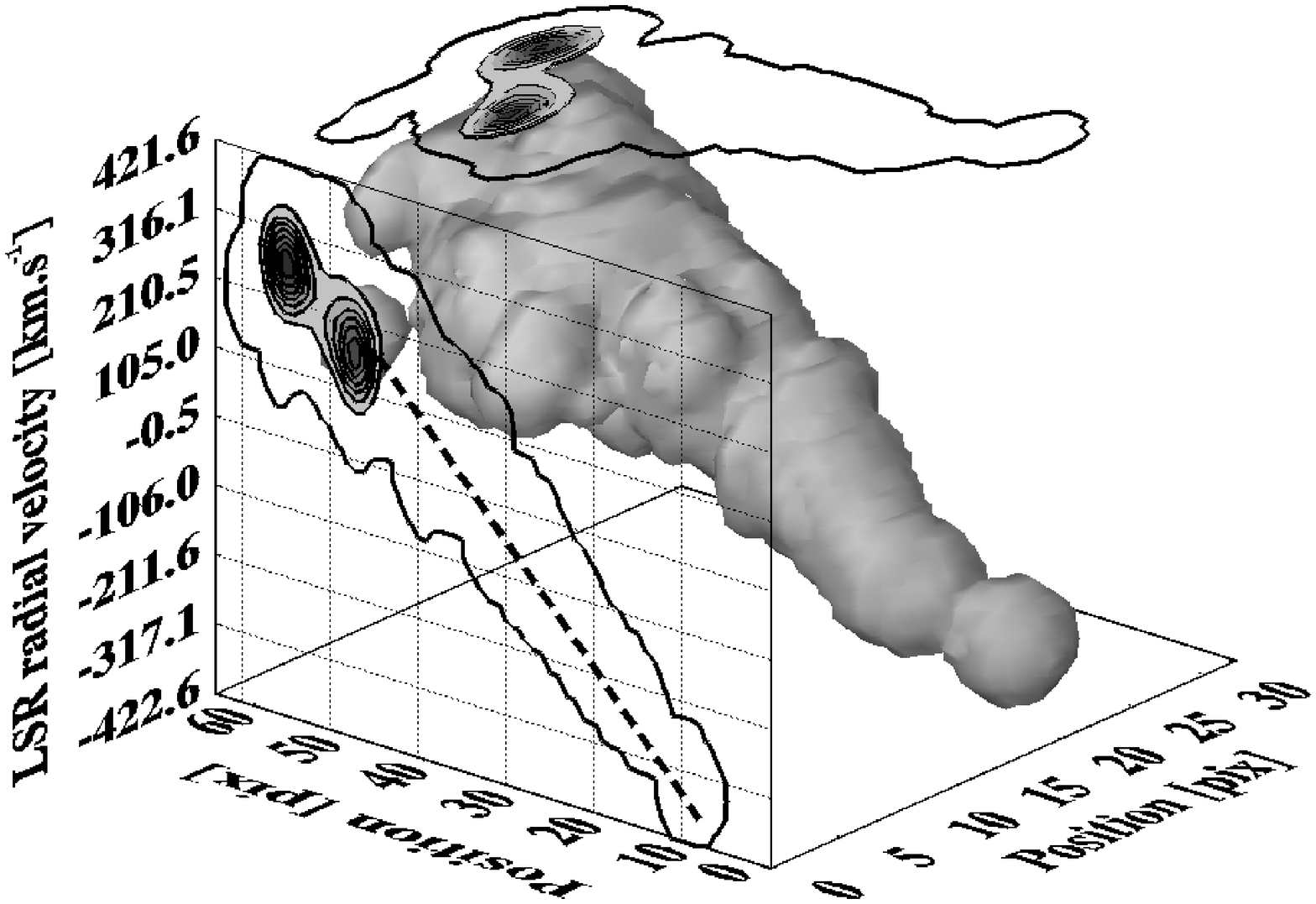}
\includegraphics[bb=30 135 475 735, angle=90, scale=.257, clip]{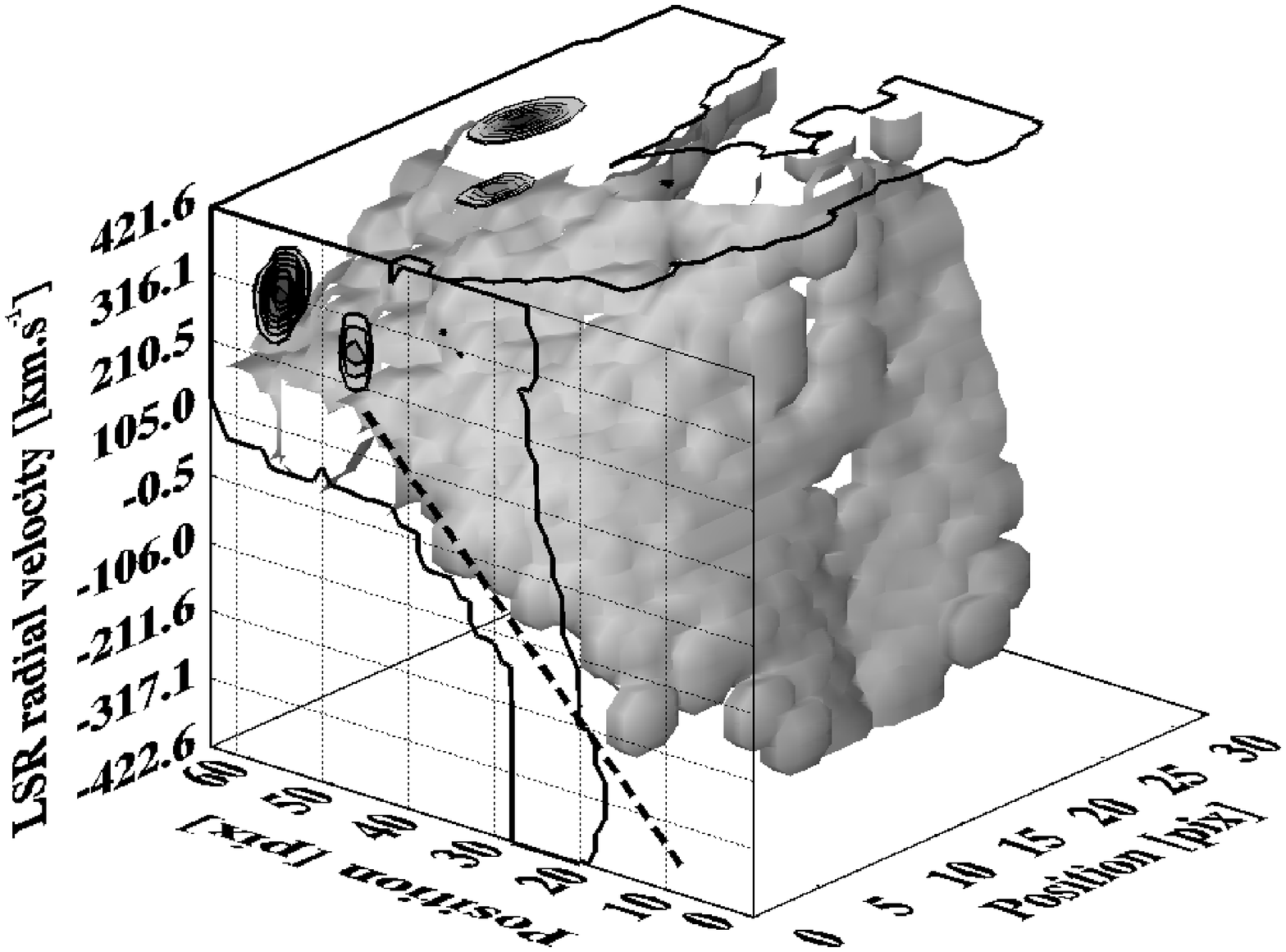}
\caption{Visualization of the entire 3\,D H\,I data--cube of the Magellanic System.
The column density isosurface $\Sigma = 10^{-5}\Sigma_\mathrm{max}$ is shown in every plot, together
with the data--cubes projected to the 2\,D maps of the integrated column density both in the position--position
and position--radial velocity spaces. 
The best model ever found in the global scale search ($F = 0.514$,
$\mu_\mathrm{lmc}^\mathrm{N}=-0.34\,\mathrm{mas\,yr^{-1}}$,
$\mu_\mathrm{lmc}^\mathrm{W}=-0.70\,\mathrm{mas\,yr^{-1}}$, left hand plot) is compared to the best model identified for
the \cite{KalliLMC, KalliSMC} data, i.e. the proper motion region delimited by Eq.~(\ref{hires_pm_1}) to~(\ref{hires_pm_4}),
($F = 0.336$,
$\mu_\mathrm{lmc}^\mathrm{N}=-1.96\,\mathrm{mas\,yr^{-1}}$,
$\mu_\mathrm{lmc}^\mathrm{W}=+0.40\,\mathrm{mas\,yr^{-1}}$, right hand plot),
and to the low--resolution compilation of the H\,I data
by \cite{Bruens05}. The dashed line in the position--velocity projections depicts the mean radial velocity gradient along the observed
Magellanic Stream.
\label{3d_maps}}
\end{figure}
}
\clearpage

\clearpage

{\onecolumn
\begin{figure}
\epsscale{1.11}
\includegraphics[angle=90,scale=.284]{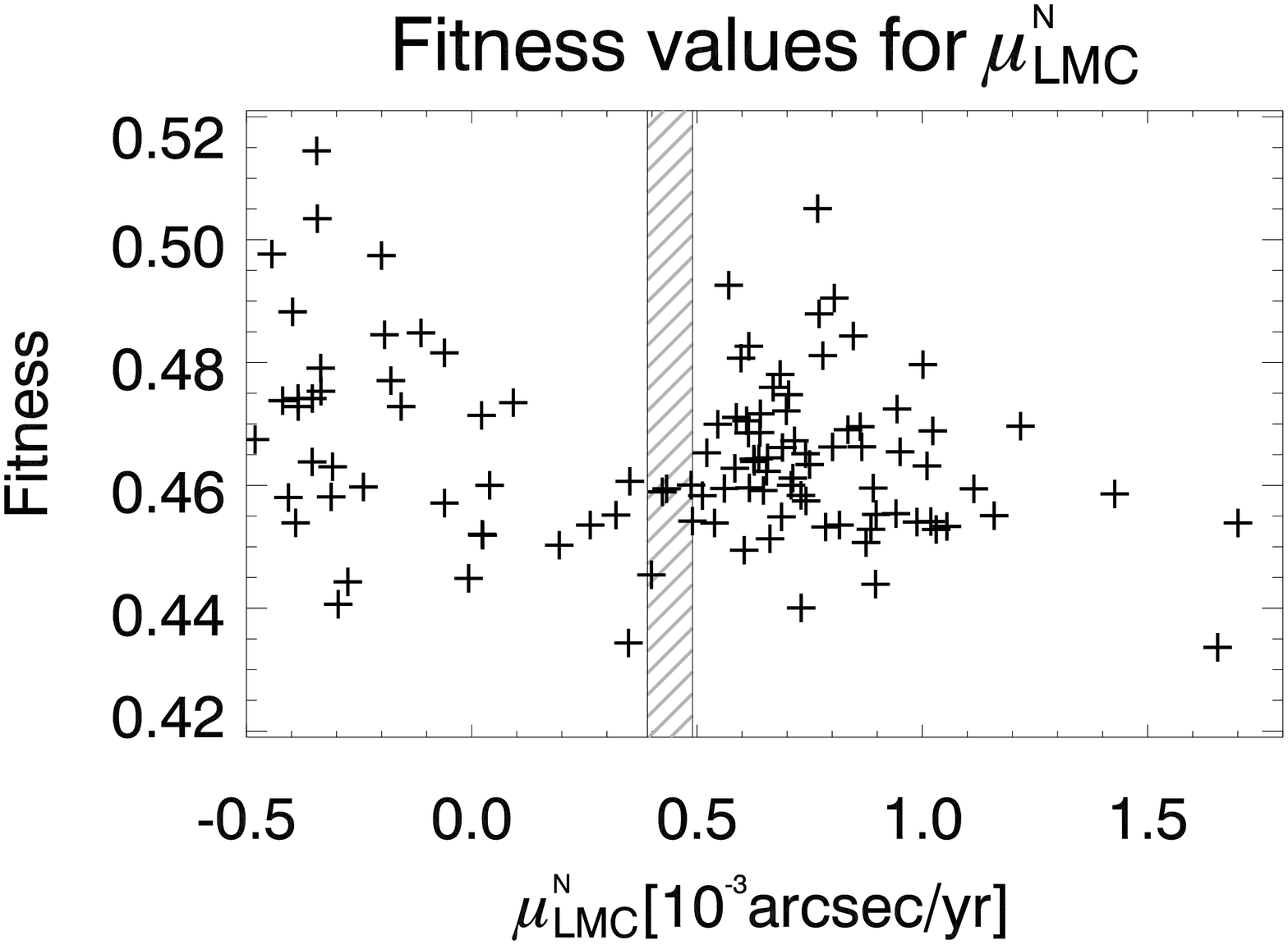}
\includegraphics[angle=90,scale=.284]{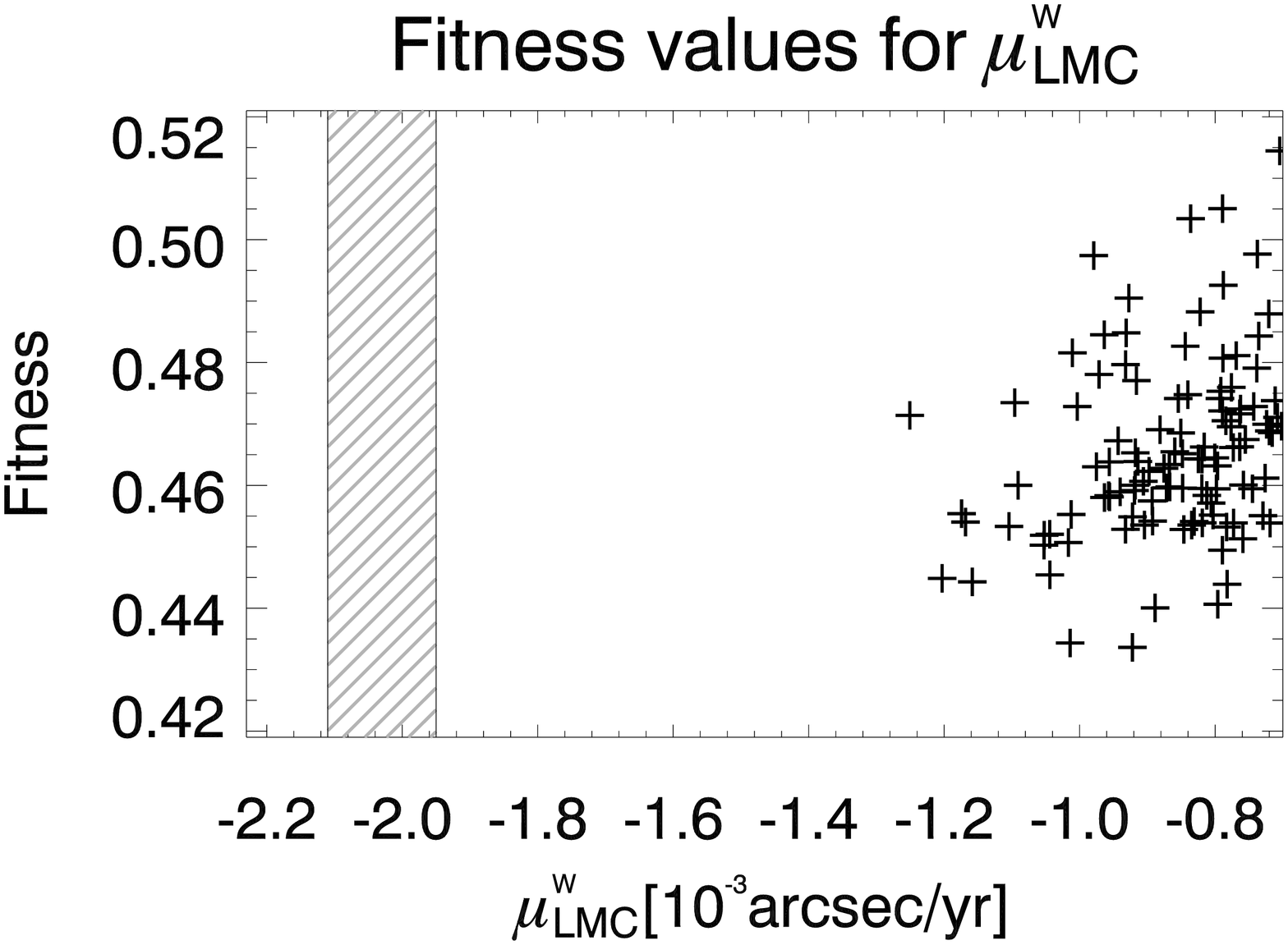} \\
\includegraphics[angle=90,scale=.284]{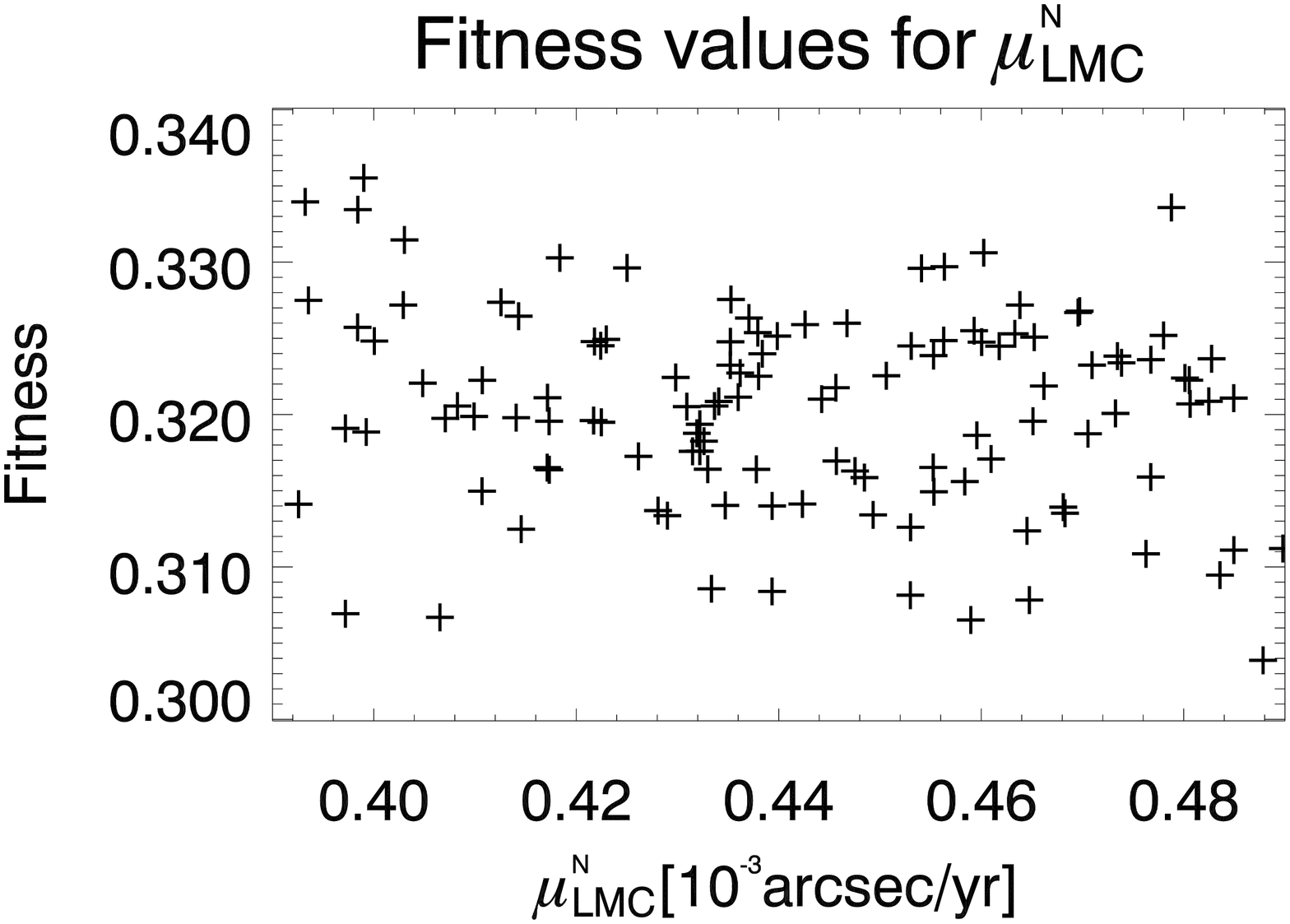}
\includegraphics[angle=90,scale=.284]{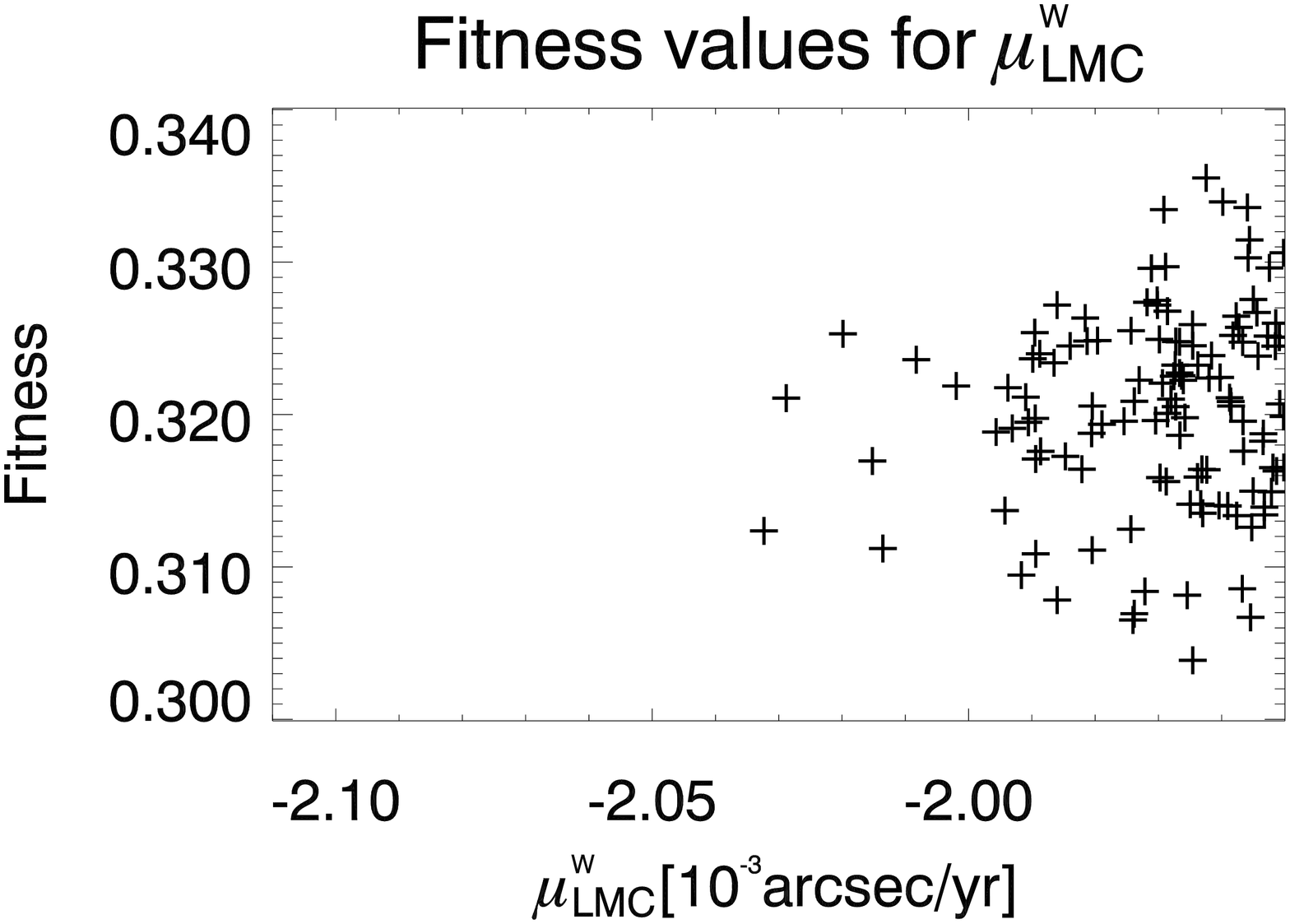}
\caption{Distribution of all GA fits of the Magellanic System over the analyzed ranges
for the LMC proper motion components \(\mu^\mathrm{N}_\mathrm{lmc}\) and
\(\mu^\mathrm{W}_\mathrm{lmc}\). The upper row presents the low--resolution search
of the original volume of the parameter space. The gray--filled areas indicate the reduced proper
motion intervals. They were studied subsequently and the resulting span of the 120 identified
GA fits is depicted in the lower row.
\label{fit-muLMC}}
\end{figure}
}
\clearpage

\end{document}